\documentclass[a4paper,12pt]{article}
\usepackage{jheppub} 

\usepackage[T1]{fontenc} 

\usepackage[percent]{overpic}
\usepackage{wrapfig}
\usepackage{tabu}
\usepackage{diagbox}

\usepackage{graphicx}
\usepackage{tikz}
\usepackage{import}
\usepackage{accents}
\usepackage{mathrsfs,amsmath,amssymb,slashed}
\usepackage{multirow,multicol}
\usepackage[percent]{overpic}
\usepackage{slashed}

\usepackage{wrapfig}
\usepackage{tabu}
\usepackage{diagbox}
\usepackage{tikz}
\usepackage{mathrsfs,amsmath,amssymb,amsthm,amsfonts,graphicx,accents,hyperref,color}
\usepackage{leftidx}
\usepackage{import}
\usetikzlibrary{decorations.pathmorphing}
\DeclareFontFamily{OT1}{rsfs}{}

\DeclareFontShape{OT1}{rsfs}{m}{n}{ <-7> rsfs5 <7-10> rsfs7 <10->rsfs10}{} 

\DeclareMathAlphabet{\mycal}{OT1}{rsfs}{m}{n}

\newcommand{\Laa}{\Lambda}

\newcommand{\B}{\mathcal{B}}

\newcommand{\ord}[1]{\mathcal{O}(#1)}
\newcommand{\laa}{\lambda}

\newcommand{\di}{\text{d}}

\newcommand{\A}{\mathcal{A}}

\newcommand{\F}{\mathcal{F}}

\newcommand{\p}{\prime}

\newcommand{\pp}{\prime\prime}

\newcommand{\Ti}{{\scriptstyle T}}
\newcommand{\Ai}{{\scriptscriptstyle A}}
\newcommand{\Bi}{{\scriptscriptstyle B}}
\newcommand{\T}{{\scriptscriptstyle T}}

\newcommand{\nn}{\nonumber}

\newcommand{\be}{\begin{equation}}
\newcommand{\ee}{\end{equation}}

\title{Asymptotic Symmetries of Maxwell Theory in Arbitrary  Dimensions at Spatial Infinity}
\author[]{Erfan Esmaeili}
\affiliation[]{\it School of Physics, Institute for Research in Fundamental
Sciences (IPM),\\ P.O.Box 19395-5531, Tehran, Iran}
\emailAdd{erfanili@ipm.ir}

\abstract{
The asymptotic symmetry analysis of Maxwell theory at spatial infinity  of Minkowski space with $d\geq 3$ is performed. We revisit the action principle in de Sitter slicing and make it well-defined by an asymptotic gauge fixing.  In consequence, the conserved charges are inferred directly by manipulating surface terms of the action. Remarkably,  the antipodal condition on de Sitter space is imposed by demanding regularity of field strength at light cone for $d\geq 4$. We also show how this condition reproduces and generalizes the parity conditions for inertial observers treated in 3+1 formulations. The expression of the charge for two limiting cases is discussed: Null infinity and inertial Minkowski observers. For the separately-treated 3d theory, a set of non-logarithmic boundary conditions at null infinity are derived by large boost limit.
}
\begin{document}
\maketitle

\section{Introduction}

In Lagrangian theories with a Lie group of global symmetries, Noether's first theorem establishes a conserved current for every generator of the corresponding Lie algebra. Noether's method, however, fails to assign conserved currents to gauge symmetries \cite{Compere:2007az}. Instead, several methods  have been proposed to associate 2-form currents $k^{\mu\nu}$ to gauge symmetries\cite{Regge:1974zd,Lee:1990nz,Anderson:1996sc,Barnich:2001jy}, which yield conserved surface charges. In gauge theories (and gravity), the asymptotic symmetry group (ASG) is the group of gauge transformations with finite surface charge\footnote{Quotiented by transformations with vanishing charge.}, and the elements are called \emph{large (or improper) gauge transformations} (in gravity, \emph{large (or improper) diffeomorphisms}).  
To obtain the asymptotic symmetry group, one fixes an appropriate gauge, and imposes certain fall-off behavior on the fields. Large gauge transformations are then the \emph{residual gauge transformations} i.e. those which preserve both the boundary conditions and the gauge.

Recent interest in ASG of Maxwell theory in flat space emanated from the discovery that soft photon theorem in QED is the Ward identity of the asymptotic symmetry group \cite{He:2014cra,Campiglia:2015qka}. 
Concerned with that motivation, research on asymptotic symmetries  is mostly performed in null slicing (Bondi coordinates) of flat space, where the surface of integration is (almost) light-like-separated from the scattering event \cite{He:2014cra, Campoleoni:2018uib,Campoleoni:2017qot, Kapec:2014zla,Satishchandran:2019pyc,Hosseinzadeh:2018dkh,Hirai:2018ijc}. The charges ``at null infinity'' have also been generalized to subleading orders \cite{Seraj:2016jxi, Campiglia:2016hvg,Campiglia:2018dyi}. 

The asymptotic symmetry group of Maxwell theory \emph{at spatial infinity} is, as far as we know, restricted to three and four dimensions \cite{Barnich:2015jua, Seraj:2016jxi, Campiglia:2017mua,Henneaux:2018gfi,Prabhu:2018gzs}. Spatial infinity examination allows  applying the canonical methods and define the ASG in a standard way. In \cite{Seraj:2016jxi}, the multipole moments of a static configuration were exhibited as asymptotic symmetry charges.  In \cite{Campiglia:2017mua}, the charges were defined in de Sitter slicing of flat space (explained later), and the null infinity charges would be recovered if the integration surface approached null infinity. We will follow much similar path in this paper, recovering \cite{Strominger:2014pwa, Satishchandran:2019pyc} at null infinity.

In this work, we study the asymptotic structure of Maxwell field in arbitrary dimensions at spatial infinity, and identify a set of boundary conditions with non-trivial ASG, generalizing previous works in four dimensions.  The ASG with our prescribed boundary conditions is local-$U(1)$ on celestial sphere $S^{d-2}$, parametrized by arbitrary functions on $S^{d-2}$. The surface charges are obtained by manipulating surface terms arising from variation of the action, circumventing standard methods. To do this, we make the action principle well-defined, by making the timelike boundary  term vanish, as done in \cite{Marolf:2000cb,Mann:2006bd,Goto:2018iay,Castro:2008ms}.  As it was shown in  \cite{Afshar:2018apx}, demanding the action principle to be well-defined determines the asymptotic gauge almost completely. This condition automatically ensures conservation of the charges for residual gauge transformations.

A key result of this paper is that we provide a rationale for imposing the antipodal matching condition in arbitrary dimensions. Previous works on gauge theories in flat space advocate a matching condition \cite{Strominger:2014pwa}  for the fields at spatial infinity $i^0$, when it is approached from future and past null boundaries $\mathcal{I^+},\,\mathcal{I}^-$. On the asymptotic de Sitter space, this condition relates the states at past and future boundaries $\mathcal{I^\pm}$. In dS/CFT studies, various antipodal conditions are proposed to make the Hilbert space well-defined \cite{Bousso:2001mw}. We show that an antipodal condition is necessary to ensure regularity of field strength at light cone for $d\geq 4$.

We will work in de Sitter slicing \cite{Ashtekar:1991vb,Prabhu:2018gzs} of Minkowski space which makes the boundary conditions manifestly Lorentz invariant. In the 3+1 Hamiltonian approach of \cite{Henneaux:2018gfi},  the formalism loses manifest Lorentz symmetry and the ASG is presented as the product of two opposite-parity subgroups. We will show how their results regarding conserved charges and parities are recovered and generalized, by focusing on specific slices of de Sitter space.

Finally, 3-dimensional theory is covered in section \ref{3d sec}. Asymptotic symmetries of 3d Einstein-Maxwell theory was studied in \cite{Barnich:2015jua} at null infinity and in \cite{Donnay:2018ckb} in near-horizon geometries. We will show by taking null infinity limit that the same set of charges (in Maxwell sector) can be obtained in a non-logarithmic expansion. In addition, our hyperbolic setup fits completely with \cite{Compere:2017knf} on  BMS$_3$ symmetry at spatial infinity. Thus, we expect that the combined hyperbolic analysis will reproduce the results of \cite{Barnich:2015jua} in its non-radiative sector.

\section{Rindler patch, action principle and conserved charges}\label{section action}
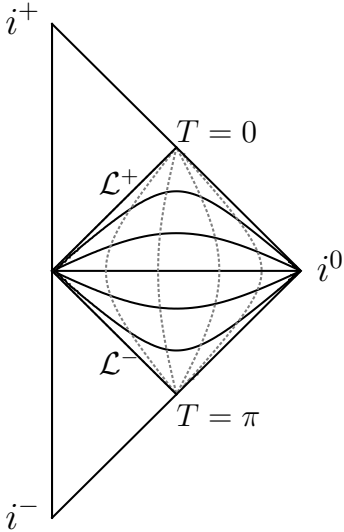
\begin{figure}[t]
 \centering
\begin{overpic}[width=0.2\textwidth,tics=1]{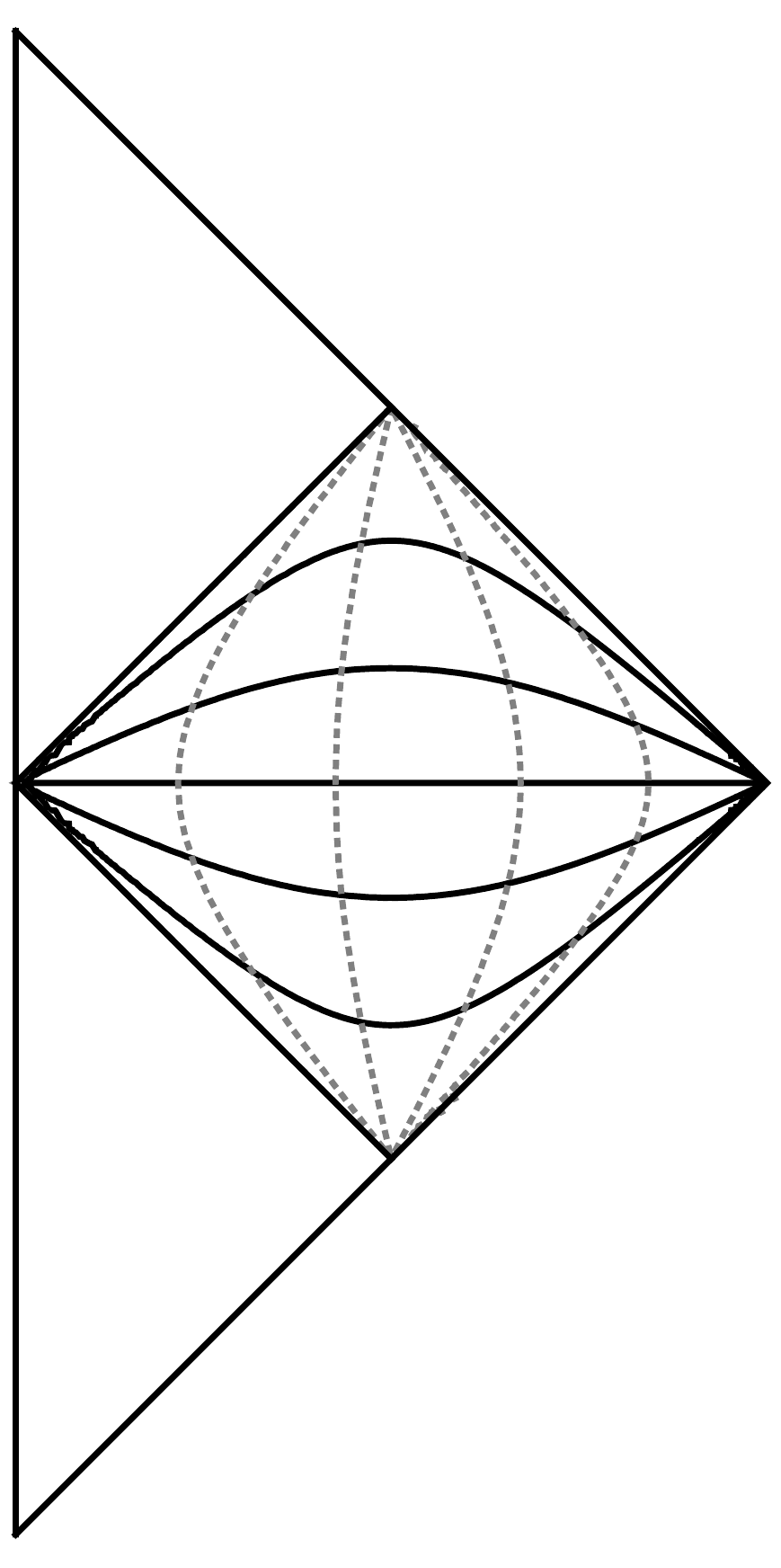}
 \put (10,30) {$\mathcal{L}^-$}
 \put (10,65) {$\mathcal{L}^+$}
 \put (25,75) {$T=0$}
  \put (25,20) {$T=\pi$}
  \put (52,48) {\large$i^0$}
\put (-8,0) {\large$\displaystyle i^-$}
\put (-8,96) {\large$\displaystyle i^+$}
 \end{overpic}
\caption{Penrose diagrams of Minkowski flat spacetime ${\cal M}_d$.  The Rindler patch covers the events outside the light cone. The solid lines are constant $T$ slices, while dotted lines are constant $\rho$ hyperboloids.}
    \label{Rind fig}
\end{figure}

Given an arbitrary point $\mathcal{O}$ in Minkowski space, one can define null coordinates $u=t-r$ and $v=t+r$. The future light cone $\mathcal{L}^+$ of $\mathcal{O}$ is the $u=0$ hypersurface, while the past light cone $\mathcal{L}^-$ is at $v=0$. $\mathcal{L}^+$ and $\mathcal{L}^-$ intersect at the origin $\mathcal{O}$. We call the set of points with space-like distance to $\mathcal{O}$, the \emph{Rindler patch} and denote it by Rind$_{d-1}$ (see figure \ref{Rind fig}). The Rindler patch is conveniently covered by coordinates $(\rho,\Ti,x^\Ai),\, A=1,\cdots ,d-2$, in which the metric is
\begin{equation}
    ds^2=\di \rho^2+\frac{\rho^2}{\sin^2\Ti}\left(-\di \Ti^2+ q_{\Ai\Bi}\di x^\Ai\di x^\Bi\right)\,,\qquad 0\leq \Ti\leq \pi\,.
\end{equation}
where 
\begin{equation}
    \left\{\begin{array}{c}
    \rho^2=r^2-t^2\\
     \cos \Ti=t/r
     \end{array}\right.
     \qquad\qquad
         \left\{\begin{array}{c}
     t=\rho\cot \Ti\\
     r=\rho/\sin \Ti
     \end{array}\right.
\end{equation}
The origin is at $\rho=0$ and undefined $\Ti$. Future  light cone $\mathcal{L}^+$ is at $(\rho=0, \Ti=0)$ and past  light cone $\mathcal{L}^-$ is at$(\rho=0, \Ti=\pi)$.
\footnote{A point at radius $r$ on $\mathcal{L}^+$ is at $(\rho=0, \Ti=0)$, by taking the limit $\Ti\to0$ such that $\rho=r\Ti$.}
Spatial infinity $i^0$ defined as the destination of spacelike geodesics is at $(\rho\to\infty, 0<\Ti<\pi)$, shown as the intersection of future and past null infinities on the Penrose diagram. The limit $(\rho\to\infty,\Ti\to0,\pi)$ covers the portion of null infinity outside the light cone.\footnote{The point at retarded time $u$ on future null infinity is reached by taking the limit $\rho\to\infty$ such that $u=-\rho\Ti/2$. Similarly, taking the limit  with fixed $v=\rho(\pi-\Ti)/2$, one arrives at the advanced time $v$ on past null infinity.}

The constant $\rho$ hypersurfaces are $(d-1)$-dimensional de Sitter spaces with radius $\rho$, invariant under Lorentz transformations about $\mathcal{O}$.  We will show de Sitter coordinates by $x^a$,  $a=2,\cdots,d$, and the unit $dS_{d-1}$ metric by $h_{ab}$.

The study is restricted to solutions with asymptotic power expansion in $\rho$
\begin{equation}\label{expand conv}
    \A(\rho,x^a)=\sum_n A^{(n)}(x^a)\rho^{-n}
\end{equation}

In some cases, we drop the superscript $(n)$ for the leading term (the least $n$) in each component to reduce clutter.

\subsection{The action principle}
\begin{figure}
	\centering


	\begin{tikzpicture}[thick,scale=0.5, every node/.style={scale=0.5}]
	


	
	\draw[magenta,thick] (8,4.3)..controls(5.5,0)..(8,-4.3)node{};
	\draw[magenta,thick] (-2,4.3)..controls(.5,0)..(-2,-4.3)node{};

	\path[fill=magenta,opacity=0.1] (-2,4.3)..controls(.5,0)..(-2,-4.3)
	arc (180:360:5cm and 1cm)
	(8,-4.3)..controls(5.5,0)..(8,4.3)  arc (0:-180:5cm and 0.8cm)
	;
	
	\draw[draw=magenta] (3,4.3) ellipse (5cm and 0.8cm);
	\draw[magenta] (-2,-4.3) arc (180:360:5cm and 1cm)node{};
	

	\draw[dashed] (-2.2,5.2)--(8.2,-5.2);
	\draw [dashed] (-2,-5.2)--(8.2,5.2);
	
	\node [rotate=65] at (-1.2,-2) {\LARGE$\rho=\rho_0$};
	\node at (3.7,0) {\LARGE$\mathcal{O}$};
	\node  at (8,0) {\LARGE$\Ti=\pi/2$};
	
	\node [rotate=63] at (8.2,3.5) {\LARGE$\Ti\to0$};
	\node [rotate=-63] at (8.2,-3.5) {\LARGE$\Ti\to\pi$};
	
	\fill [fill=red!50!white,very thick,opacity=.8]plot  coordinates{(3,0) (8,4.2)(7.8,4.1)(7,3.8)(6,3.65)(3,3.45)(0,3.65)(-1,3.8)(-1.8,4.1)(-2,4.2)(3,0)};
		\fill [fill=red!50!white,very thick,opacity=.8]plot  coordinates{(3,0) (8,-4.3)(7.8,-4.6)(7,-4.9)(6,-5.1)(3,-5.3)(0,-5.1)(-1,-4.9)(-1.8,-4.6)(-2,-4.3)(3,0)};
	
	\node [] at (3,2) { \LARGE$I_2$};
		\node []at (3,-3) {\LARGE $I_1$};
			\node[	] at (1,0) {\LARGE	 $B$};
	\end{tikzpicture}

	\caption{The region where we define the action problem. It is confined by  initial and final cones $I_1$ and $I_2$ (\emph{e.g.} at constant $T$), intersecting at $\mathcal{O}$. The region is not bounded in $\rho$ direction, so $I_{1,2}$ are Cauchy surfaces where initial and final data are fixed. The boundary terms are computed at constant-$\rho$ hyperboloids ($B$). Dashed lines show the light-cone.}
	\label{action fig}
\end{figure}
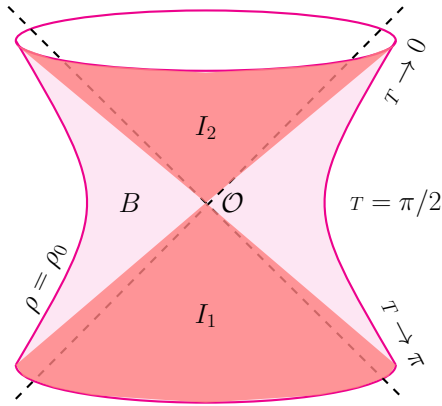

In the Lagrangian formulation of physical theories, the \emph{classical trajectories} of the dynamical variables $\Phi^i$ are stationary points of an action functional

\begin{equation}
  \frac{\delta S}{\delta \Phi^i}\Big|_{\Phi^i_{cl.}}=0\,
\end{equation}
for fixed initial and final values. In field theories, 
the functional derivative of the action is well-defined, if variation of dynamical fields leaves no boundary terms. In our setup, there are two spacelike boundaries $I_{1,2}$ and one timelike boundary $B$ lying on asymptotic de Sitter space (see figure \ref{action fig}). Data on spacelike boundaries is fixed, so we must ensure that the boundary term on $B$ either vanishes or itself is a total derivative.

For Maxwell theory with action\footnote{Notation: $g=|\text{det}g_{\mu\nu}|$ for all metrics involved.}
\begin{equation}\label{action}
S=\int_{M_d} \sqrt{g}\left(-\frac{1}{4}\F_{\mu\nu}\F^{\mu\nu}+\A_\mu J^\mu\right)\,,
\end{equation}
the timelike boundary term  is
\begin{equation}
\int_B \rho^{d-1}\sqrt{h}\delta \A_a\F^{a\rho}\label{maxwell boundary}
\end{equation}
We will show that for specific boundary conditions and an asymptotic gauge fixing, the boundary term does vanish.

	\subsection{Conserved charges}
For the specific example of Maxwell theory, we show that with a well-defined action principle at hand, one can define  conserved charges for gauge transformations of the theory, and identify the asymptotic symmetry group as the group of gauge transformations having  finite charge. 

Consider variation of the action around a solution to equations of motion\footnote{Notation: $\approx$ is equality when equations of motion hold.}
\begin{equation}
    \delta S[\Phi]\approx \int_{I_2}\mathcal{I}(\Phi,\delta\Phi)-\int_{I_1}\mathcal{I}(\Phi,\delta\Phi)+
    \int_{B}\mathcal{B}(\Phi,\delta\Phi)\label{variation of action}
\end{equation}
If the field variation is a gauge transformation (or a diffeormorphism in  gravity theories), then, the $\mathcal{I}$ integrands in \eqref{variation of action} become total derivatives, so the first two terms becomes codimension-2 integrals on $\partial I_1$ and $\partial I_2$. This can be checked in specific examples, and a proof for gravity case is given in \cite{Iyer:1994ys}

If the action principle is well-defined, the $B$-integral on the timelike boundary is either vanishing, or a total derivative on the hyperboloid (so that it becomes a surface integral on boundaries of $B$). As a result, the gauge transformation of the action becomes the difference of two codimension-2 integrals on shell
\begin{equation}\label{variation}
    \delta_\laa S\approx \int_{\partial I_2}\mathcal{C}(\Phi,\delta_\laa\Phi)-\int_{\partial I_1}\mathcal{C}(\Phi,\delta_\laa\Phi)\,.
\end{equation}
The left-hand-side depends on the the explicit form of the action.  If the action is gauge invariant ($\delta_\laa S=0$), \eqref{variation} shows that the integral $\int_{\partial I}\mathcal{C}$ is independent of the surface of integration; thus we can identify the codimension-2 integrals as the conserved charges corresponding to the gauge transformation $\delta_\laa$.

\subsection*{Covariant phase space method}
Let us compare the procedure above with covariant phase space method. The symplectic form of the theory is nothing but variation of the action surface terms
\begin{equation}
    \Omega=\int_I \mathcal{I}(\delta\Phi,\delta^\prime\Phi)
\end{equation}
for two field variations $\delta,\delta^\p$, and $\mathcal{I}$ is defined in \eqref{variation of action}. Taking a second variation of \eqref{variation of action} shows that in general $\Omega$ is not conserved since its flux at timelike boundary $B$ is non-vanishing and given by
\begin{equation}\label{flux}
      \int_{B}\mathcal{B}(\delta\Phi,\delta^\prime\Phi)
\end{equation}
where the integrand $\B$ is again defined in \eqref{variation of action}. Therefore, eliminating the symplectic flux is equivalent to making the action principle well-defined. For the conservation of the symplectic form, the flux \eqref{flux} need not be strictly vanishing. It is enough, if possible, to make it a total divergence reducing the expression to codimension-2 integrals on $\partial B$:
\begin{equation}
   \int_{B}\mathcal{B}(\delta\Phi,\delta^\prime\Phi)\equiv-\Omega^{\text{\tiny b'dry}}\big|^{\partial I_2}_{\partial I_1}
\end{equation}
Finally, $\Omega^{\text{\tiny b'dry}}$ can be added to $\Omega$ as a surface term, leading to conserved charges. This subtraction is a \emph{Y ambiguity} in covariant phase space terminology \cite{Lee:1990nz}. This procedure was done in \cite{Campiglia:2017mua} for 4d Maxwell theory. It can be readily generalized to arbitrary dimensions by appropriate choice of boundary conditions. However, we decide to bypass the symplectic form by working directly with the action.
\subsection{Field equations in de Sitter slicing}\label{subs equation}
Written in coordinates $(\rho,x^a)$, the field equations and Bianchi identities are
\begin{subequations}
\begin{align}
    D_a\F^{a\rho}&=J^\rho\label{eq rho}\\
    \frac{1}{\rho^{d-1}}\partial_\rho(\rho^{d-1}\F^{\rho a})+D_b\F^{ba}&=J^a\label{eq a}\\
    \partial_\rho\F_{ab}+2\partial_{[a}\F_{b]\rho}&=0\label{Bianchi rho}\\
   \partial_{[a}\F_{bc]}&=0 \label{Bianchi a}
\end{align}
\end{subequations}
where $D$ is the covariant derivative on dS$_{d-1}$. Analyzing the solutions suggests appropriate boundary conditions for the theory.
Note that $\F_{a\rho}$ and $\F_{ab}$ are distinct Lorentz invariant components. First we ask if there are solutions to equations of motion once either of them is set to zero.
\begin{enumerate}
    \item 
If we set $\F_{ab}=0$, by \eqref{eq a}  we have
\footnote{Notation: $\sim\ord{\rho^n}$ means all powers not exceeding $n$, while $\propto \rho^n$ means the $n$-th power of $\rho$ exclusuvely.}
$ \F^{a\rho}\propto\rho^{1-d}$ .
Furthermore, by \eqref{Bianchi rho} and \eqref{eq rho},
\begin{equation}
    \F_{a\rho}=\partial_a\Psi=\rho^{3-d}\partial_a\psi(x^b)\,,\qquad   D^aD_a\Psi=J^\rho \label{indp electric}
\end{equation}
In this case, the solution consists of a scalar degree of freedom $\psi$.
\item  In general,
 $\F_{ab}$ is closed on de Sitter space by \eqref{Bianchi a}, thus it is locally exact $\F_{ab}=(\di\A)_{ab}$.
  Switching $\F_{a\rho}$ off, fixes the $\rho$-dependence by \eqref{Bianchi rho} to $ \F_{ab}\propto\rho^0$.
 Finally, the field equation \eqref{eq a} reduces to
\begin{equation}
    D^aD_{[a}A^{(0)}_{b]}=0 \label{indp magnet}
\end{equation}
(Notation is explained in \eqref{expand conv}).
\end{enumerate}

Any other solution involves both $\F_{ab}$ and $\F_{a\rho}$. The solutions with power-law fall-off in $\rho$ correspond to multipoles of electric and magnetic branes. Electric monopoles generate the independent solution \eqref{indp electric} for $\F_{a\rho}$, while magnetic mono-poles(-branes) generate the independent solution \eqref{indp magnet}  for$\F_{ab}$. Their multi-poles generate fields of lower fall-off which mix $\F_{a\rho}$ and $\F_{ab}$. On the contrary, arranging monopoles to build lines of charge will generate stronger fields at infinity, but in any case mix $\F_{a\rho}$ and $\F_{ab}$.
\footnote{
For example, for an electric dipole, $\F_{\rho a}\propto \rho^{2-d}$, and by Bianchi identity \eqref{Bianchi rho}, $\F_{ab}\propto\rho^{3-d}$. Define $\psi_a\equiv F^{(d-2)}_{a\rho}$, Then,
\begin{subequations}\label{dipole eq}
\begin{align}
  (d-3)  F^{(d-3)}_{ab}=(\di\psi)_{ab},\qquad
    D^b(\di\psi)_{ba}-(d-3)\psi_a=0
\end{align}
away from sources.
\end{subequations}

}

    Denote the set of solutions for electric monopoles given in \eqref{indp electric} by $\mathcal{E}$. This space covers moving electric charges in space, which are passing the origin simultaneously at  $t=0$, hence their worldlines cross $\mathcal{O}$. The field strength is $\F_{\rho a}\propto \rho^{3-d}$ with no subleading terms. For an arbitrary configuration of freely moving charges, the leading component of asymptotic field is an element of $\mathcal{E}$, but subleading terms are generally present. In other words, the definition of $\mathcal{E}$ is Lorentz invariant, but not Poincar\'e  invariant. 
\emph{$\mathcal{E}$ encodes the information of charge values $q_n$ and their velocities $\vec{\beta}_n$.}
The space $\mathcal{E}$ is isomorphic to the space of boost vectors $\vec{\beta}$, that is $\mathbb{R}^{d-1}$. The space of conserved electric charges we will construct is also isomorphic to $\mathbb{R}^{d-1}$; each point of this space with coordinate vector $\vec{\beta}$ is a conserved charge and gives the total electric charge in space, moving with that specific boost.

The set of solutions \eqref{indp magnet} covers magnetic monopoles moving freely in space and crossing the origin at $t=0$. In dimensions larger than 4,  the magnetic monopoles are replaced by extended magnetic branes since the dual field strength $\ast F$ is a $(d-2)$-form in that case. We are considering boundary conditions which exclude magnetic charges in this work.

\section{Four and higher dimensions}
In this section, we exploit the asymptotic symmetries of Maxwell theory in dimensions higher than three. First, we present a set of well-motivated boundary conditions on field strength tensor. Nonetheless, existence of large gauge transformations demand that the gauge field be finite at infinity.  That will necessitate an asymptotic gauge choice to make the action principle well-defined. Finally, we find the conserved charges of the theory at spatial infinity by computing the on-shell action.

\subsection{Boundary conditions and the action principle}
The electromagnetic field of a static electric charge is
\footnote{
For a spherically symmetric field we have
\begin{equation}
Q=\int_{S^{d-2}}\sqrt{q}r^{d-2} \F_{tr}\quad \rightarrow\quad 	\F_{tr}=\frac{Q}{a_{d-2}}r^{2-d}\nn
\end{equation}
where $a_{d-2}$ is the area of a $(d-2)$-sphere. In hyperbolic coordinates we have
\begin{equation}
\F_{\T\rho}=-\frac{\rho}{\sin\Ti}\F_{tr}=-\frac{Q}{a_{d-2}}\rho^{3-d}\sin^{d-3}\Ti\nn
\end{equation}
}
  $\F_{\T\rho}\propto\rho^{3-d}$. Applying a boost (which belongs to the isometry group of the hyperboloid) will turn on other de Sitter components of $\F_{a\rho}$ with the same fall-off; so one generally has $\F_{a\rho}\propto\rho^{3-d}$. Therefore, we propose the following boundary conditions for $d$-dimensional theory
\begin{equation}
    \F_{\rho a}\sim\ord{\rho^{3-d}}\,,\qquad \F_{ab}\sim \ord{\rho^{3-d}}\,.\label{maxwell b.c.}
\end{equation}
The $\F_{ab}$ components arise because of electric multipoles (\emph{c.f.} \S \ref{subs equation}).
The leading component of $\F_{a\rho}$ is in $\mathcal{E}$ of $\S$ \ref{subs equation} and satisfies
\begin{equation}
    F^{(d-3)}_{a\rho}=\partial_a\psi\,,\qquad D_aD^a\psi=j^{(1-d)}_\rho
\end{equation}


Components of gauge field that saturate \eqref{maxwell b.c.} behave like
\begin{equation}
 \A_{a}\sim\ord{\rho^{3-d}}\qquad  \A_{\rho}\sim\ord{\rho^{3-d}}
\end{equation}
. Plugging into  \eqref{maxwell boundary}, the boundary term falls like  $\ord{\rho^{3-d}}$. For $d>3	$, the action principle is well-defined. However, this choice will make the charges for all gauge transformations vanish. For instance, the Gauss law
\begin{equation}
Q=\int_{S^{d-2}}\ast\F
\end{equation}
is regarded as the charge for gauge transformation with $\laa=1$, which is excluded if $\A_{a}\sim\ord{\rho^{3-d}}$ in dimensions higher than three. The theory enjoys non-trivial ASG, only if $\delta\A_a\sim\ord{1}$. Thus, our prescribed boundary condition is as follows:  $\A_a\sim\ord{1}$ but the first few terms in the asymptotic expansion of $\A_a$ are pure gauge\footnote{By ``pure gauge'' we mean a flat connection; a configuration gauge equivalent to $\A_\mu=0$, although it may involve an improper gauge transformation (that with non-zero charge).}, such that $\F_{ab}\sim \ord{\rho^{3-d}}$.
Previous works in four dimensional Maxwell theory allow magnetic monopoles. That would make $\F_{ab}\sim\ord{1}$ so the leading term of the gauge field would not be pure gauge. Here we are not taking  account of magnetic charges though.

 With the aforementioned boundary condition, the boundary term of the action will be finite
\begin{equation}
    \int_B \sqrt{h}\,\delta A^{(0)}_bF_{(d-3)}^{b\rho}=
     \int_B \sqrt{h}\,\delta A^{(0)}_b\partial^b\psi
\end{equation}
According to \eqref{maxwell b.c.}, $F_{ab}^{(0)}=0$ so the leading term is (locally) pure gauge $A^{(0)}_b=\partial_b\phi$. Consequently, after integration by parts, the boundary term of the action vanishes \emph{on shell}, by equation of motion $D^2\psi=0$ (up to a total divergence on $B$). However, we request \emph{off-shell} vanishing of the boundary term, since the variational principle must entail the equations of motion, and they can not be used \emph{a priori}.

One way out is to fix the asymptotic  gauge $\delta D^aA_a^{(0)}=0$, for which the boundary term becomes a total divergence on $B$ after an integration by parts. There are also other possibilities.
The Lorenz gauge at leading order is
\begin{equation}
    D^aA_a^{(0)}+\alpha(d-2)A_\rho^{(1)}=0\,,\qquad \alpha=1\label{Lorenz gauge}
\end{equation}
and by our boundary conditions on field strength, $A_{\rho}^{(1)}=0$ for $d>4$.  Thus, the Lorenz gauge, or its extension to general $\alpha$ will make the action principle well-defined in dimensions strictly higher than 4. In four spacetime dimensions, $A_{\rho}^{(1)}=\psi$ (up to a constant number which drops from derivatives) so it is necessary to add a boundary term
\begin{equation}
S_b=-\alpha\int_{B_3}\sqrt{h}\psi^2\qquad\qquad \text{for} \,\,d=4
\end{equation}
to make the action well-defined\cite{Afshar:2018apx}.
\subsection{Conserved charges from action}
The action with a solution to equations of motion plugged in, is a functional of initial and final field values (or boundary values in Euclidean versions); That is how classical trajectories are defined. For Maxwell theory,
\begin{equation}\label{maxwell onshell action}
S\approx\int_{I_2}\sqrt{\gamma}n_\T\,\A_{\mu}\F^{\mu\T}-\int_{I_1}\sqrt{\gamma}n_\T\A_{\mu}\F^{\mu\T}+\int_B\sqrt{h}\, \partial_a\phi\partial^a\psi
\end{equation}
$\gamma$ is the induced metric on $I$ and $n^\mu$ is its future-directed normal vector.
Varying \eqref{maxwell onshell action} by gauge transformations $\delta\A_{\mu}=\partial_\mu\Laa$, and using field equations following an integration by parts gives\footnote{In equation \eqref{maxwell onshell action 2}, the induced metric on $\partial I$ yields a determinant factor $\rho^{d-2}\sin^{2-d}\Ti$. On the other hand, $n_\T=-\frac{\rho}{\sin\Ti}(0,1,\vec{0})$. The combination is equal to $-\sqrt{g}$.}
\begin{equation}\label{maxwell onshell action 2}
\delta_\Laa S-\int_{I_2}\Laa J^\T+\int_{I_1}\Laa J^\T\approx-\int_{\partial I_2}\sqrt{g}\,\laa \F^{\rho \T}+\int_{\partial I_1}\sqrt{g}\,\laa \F^{\rho \T}+\int_B \sqrt{h}\,\partial_a \laa\partial^a\psi\,.
\end{equation}
where $\laa=\Laa^{(0)}$. The explicit form of Maxwell action \eqref{action} shows that the left-hand-side above  is the flux through spatial boundary:
\begin{equation}
\delta_\Laa S-\int_{I_2}\Laa J^\T+\int_{I_1}\Laa J^\T=\int_B\sqrt{h}\rho^{d-1}\laa J^\rho
\end{equation}
We can make this ``charge flux'' vanish asymptotically by
 the additional assumption $J^\rho\sim \ord{\rho^{-d}}$. This condition ensures that the system is localized and the charges are conserved.  So far we made the left-hand-side in \eqref{maxwell onshell action 2} vanish; let us look at the other side.

 Recall that the action principle necessitated  fixing the asymptotic Lorenz gauge \eqref{Lorenz gauge}, leaving residual gauge transformations 
\begin{equation}
\delta A_a^{(0)}=\partial_a\laa \,,\qquad D^aD_a\laa=0\,,
\end{equation}
with arbitrary subleading terms. The condition on $\laa$ allows us to turn the very last term in the right-hand-side of \eqref{maxwell onshell action 2} into a total divergence on $B$. As a result we manage to prove that the quantity
\begin{equation}\label{charge maxwell 4}
Q_\laa=\int_{\partial I}\sqrt{g}\left(\laa \F^{ \T\rho}-\partial^{\T}\laa\psi\right)
\end{equation}
is independent of $I$; i.e. conserved.

\subsection{Light cone regularity and antipodal identification}
$\laa$ and $\psi$ both satisfy
\begin{equation}
    D^aD_a f(x^b)=0 \label{maxwel de sitter eom}
\end{equation}
and the solution is obtained by spectral decomposition of Laplace operator on $S^{d-2}$, being $\mathcal{D}^2Y_\ell(\hat{x})=-\ell(\ell+d-3)Y_\ell(\hat{x})$. Then, \eqref{maxwel de sitter eom} will
simplify to
\begin{equation}
    (1-y^2)f^{\pp}_\ell(y)+(d-4)yf_\ell^\p(y)+\ell(\ell+d-3)f_\ell(y)=0\qquad y=\cos\Ti\,.
\end{equation}
The general solution is
\begin{equation}\label{solution}
    f(y,\hat{x})=(1-y^2)^{\frac{d-2}{4}}\sum_{\ell=1}Y_\ell(\hat{x})\left(a_\ell P_{(2l+d-4)/2}^{(d-2)/2}(y)+b_\ell Q_{(2l+d-4)/2}^{(d-2)/2}(y)\right)\,,
\end{equation}
where $P_l^m$ and $Q_l^m$ are associated Legendre functions of the first and second kind respectively. For $\ell=0$, the solutions are
\begin{equation}
    a_0+b_0 y\,\,{}_2F_1(\frac{1}{2},\frac{4-d}{2},\frac{3}{2},y^2)\,.
\end{equation}

As far as field equations are concerned, the whole set of solutions in \eqref{solution} with two sets of coefficients are admissible both for $\psi$ and $\laa$. In previous works in four dimensional Maxwell theory, a boundary condition, the \emph{antipodal matching condition} \cite{Strominger:2014pwa}, was imposed such that one of branch of the solutions in \eqref{solution} was allowed for $\psi$ and the other for $\laa$. Here we will provide a rationale for the antipodal matching condition in higher dimensions. 

The field strength tensor $\F$ being a physical field must be regular at light cone $\mathcal{L}^\pm$ (i.e. $u=0$ and $v=0$ surfaces in advanced/retarded Bondi coordinates). Recall that in $\mathcal{E}$ space, $\F_{a\rho}=\rho^{3-d}\partial_a\psi$ in $d$ dimensions, which diverges at $\rho=0$ in dimensions larger than three. Near $\mathcal{L}^+$ (located at $\rho=0, \Ti=0$), $\psi$ must decay at least like $ \Ti^{d-2}$, to make $\F_{\T\rho}$ finite. 

The light cone behavior of solutions \eqref{solution} is\footnote{In four dimensions, the subleading term for $f_+$ is $\ord{ \Ti^2\log { \Ti}}$}
\begin{equation}
f_-= {\Ti}^{d-2}\bar{\psi}(\hat{x})+\ord{ \Ti^{d}}\,,\qquad
f_+=\bar{ \laa}(\hat{x})+\ord{\Ti^2}\,.
\end{equation}
Request for light cone regularity leads us to take $f_-$ for $\psi$ as a boundary condition on $\mathcal{L}^+$, hence the notation $\bar{ \psi}$. Similar argument can be made at $\mathcal{L}^-$ at $\Ti\to\pi$. Extracting $f_-$ from \eqref{solution} amounts to setting $b_\ell=0$ in even dimensions, and setting $a_\ell=0$ in odd dimensions (and keeping $b_0$ in all dimensions). These conditions can be summarized as \emph{antipodal identification of solutions on $dS_{d-1}$}
\begin{equation}\label{parity cond1}
    \psi(\Ti,\hat{x})=-\psi(\pi-\Ti,-\hat{x})\,.
\end{equation}
This is a well-known condition in dS/CFT studies \cite{Strominger:2001pn}. Gauge parameters with non-vanishing charge \eqref{charge maxwell 4} must reside in $f_+$ set. These are even under de Sitter antipodal map
\begin{equation}\label{parity cond2}
   \laa(\Ti,\hat{x})=\laa(\pi-\Ti,-\hat{x})\,.
\end{equation}

Note that the conditions \eqref{parity cond1} and \eqref{parity cond2} hold on the entire de Sitter space and in particular for $\Ti=0$, relating the fields on future and past boundaries of the hyperboloid
\begin{align}
\psi(0,\hat{x})&=-\psi(\pi,-\hat{x})\,\\
 \laa(0,\hat{x})&=\laa(\pi,-\hat{x})
\end{align}
The fields on left-hand-side live on the past of future null infinity $\mathcal{I}^+_-$ while those on right-hand-side live on the future of past null infinity $\mathcal{I}^-_+$. Therefore $\laa$ and $F_{a\rho}=\partial_a\psi$ are both even under antipodal map between future and past null infinity.
\subsection{Charge at null infinity}
In the Rindler patch, one can approach the light cone hypersurface $\mathcal{L}^+\cup\mathcal{L}^-$ from \emph{outside}. The charge \eqref{charge maxwell 4} takes a simpler form in that limit: The second term in \eqref{charge maxwell 4} vanishes, while the first term becomes
\begin{equation}
Q_\laa=-(d-2)\int_{S^{d-2}}\sqrt{q}\bar{ \laa}\bar{\psi}
\end{equation}
 The leading field strength at null infinity becomes 
\begin{equation}\label{null charge}
    \F_{ur}=-\frac{1}{r}\F_{\T\rho}=(d-2)r^{2-d}\bar{\psi}(\hat{x})+\ord{r^{1-d}}
\end{equation}
Hence, the familiar expression for surface charges at future null infinity is recovered
\begin{equation}
Q_\laa=\int _{S^{d-2}}\sqrt{g}\,\bar{ \laa}\,\F_{ur}
\end{equation}
\subsection{Inertial observers}
Consider a Minkowski observer with coordinates $(t,r,x^\Ai)$, who advocates a ``3+1 formulation'' of $d$-dimensional theory. Boundary conditions restrict Cauchy data residing in constant-time hypersurfaces at large $r$. It is implicitly presumed that time interval $\Delta t$ between Cauchy surfaces is much smaller that the radius $r$ beyond which is conceived as ``asymptotic region''. This $\Delta t/r\to 0$ condition makes all Cauchy surfaces to converge at $\Ti=\pi/2$ ``throat'' on the asymptotic de Sitter space. Infinitesimal Lorentz boosts will incline this surface, though, to $\Ti=\vec{\beta}\cdot\hat{x}+\pi/2$.

The solutions to (second order) equations of motion on dS$_{d-1}$ are specified by initial/final data on past/future boundaries of de Sitter space $\mathcal{I}^-_+/\mathcal{I}^+_-$. When an additional \emph{antipodal condition} is imposed, only one set of data on either boundary suffices (and the other one is determined by e.o.m.). When the spacetime is restricted to a cylinder around $\Ti=\pi/2$, the solution can be specified by a couple of \emph{independent data} $\Phi$ and $\partial_\T\Phi$ (and higher time derivatives determined by e.o.m.). The antipodal condition  then halves the possibilities in each one by a restriction on \emph{angular dependence}, as explained below.

 Here, we would like to focus around $\Ti=\pi/2$ surface and translate previous results to a canonical language. First of all, the coordinates are related as
\begin{equation}
t=\rho\cot \Ti\cong \rho (\frac{\pi}{2}-\Ti)\,,\qquad\qquad r=\frac{\rho}{\sin \Ti}\cong \rho\,.
\end{equation}

Next, recall that $A_a^{(0)}=\partial_a\phi$, which implies that
\begin{equation}
A_t(\hat{x})=\frac{1}{r}\partial_{\T}\phi(\frac{\pi}{2},\hat{x})\,,\qquad A_\Bi(\hat{x})=\partial_\Bi\phi(\frac{\pi}{2},\hat{x})
	\end{equation}
In four dimensions, $A_t$ receives an additional contribution $-\psi(\frac{\pi}{2},\hat{x})/r$. The radial components may be written as
\begin{equation}
A_r=\partial_r\Laa+r^{3-d}\bar{A}_r(\hat{x})+\ord{r^{2-d}}\,,\qquad\Laa\sim\ord{r^0}\,.
\end{equation}
where
\begin{align}
\bar{A}_r(\hat{x})&=\psi(\frac{\pi}{2},\hat{x})\,
\end{align}
The field strength is given by
\begin{align}
\pi^r\equiv\sqrt{g}F^{rt}&=-\sqrt{q}\,\partial_\T\psi(\frac{\pi}{2},\hat{x})\,,\qquad \pi^B\equiv\sqrt{g}F^{Bt}\sim\ord{r^{-2}}
\end{align}
The ``momenta'' $\pi^i$ are symbolic in this discussion, but they are equal to momenta in a true Hamiltonian formulation. Finally, the gauge parameter divides into
\begin{equation}
\laa(\hat{x})\equiv\laa(\frac{\pi}{2},\hat{x})\,,\qquad\mu(\hat{x})\equiv\partial_\T\laa(\frac{\pi}{2},\hat{x})\,.
\end{equation}
The antipodal conditions \eqref{parity cond1} and \eqref{parity cond2} imply
\begin{equation}
\bar{A}_r(\hat{x})=-\bar{A}_r(-\hat{x})\qquad \pi^r(\hat{x})=+\pi^r(-\hat{x})\qquad
A_\Bi(\hat{x})=-A_\Bi(-\hat{x})
\end{equation}
and\footnote{Parity of $\pi^B$ can not be inferred from leading fields. For electric dipoles, $\pi^B(-\hat{x})=+\pi^B(\hat{x})$.}
\begin{equation}
\qquad \mu(\hat{x})=-\mu(-\hat{x})\qquad \laa(\hat{x})=+\laa(-\hat{x})
\end{equation}
In even spacetime dimensions, these are \emph{parity conditions}, cause the antipodal map $\hat{x}\to-\hat{x}$ reverses the orientation of $S^{d-2}$ (the volume form shifts sign). In odd dimensions, however, the map is a rotation about the origin, preserving the orientation. These conditions are preserved under boosts. The connected part of Lorentz group $SL(d-1,1)$, commutes with parity and time-reversal, thus the antipodal conditions \eqref{parity cond1} and \eqref{parity cond2} hold in any Lorentz frame. Explicitly, for an infinitesimally boosted frame and keeping the terms at zeroth order of $\Ti$ we have
\begin{equation}
\psi^\p(\Ti^\p=\frac{\pi}{2},-\hat{x}^\p)=\psi(\frac{\pi}{2}-\vec{\beta}\cdot\hat{x},-\hat{x})=-\psi(\frac{\pi}{2}-\vec{\beta}\cdot\hat{x},\hat{x})=-\psi^\p(\Ti^\p=\frac{\pi}{2},\hat{x}^\p)
\end{equation}
In the second equality we have used the antipodal conditions and the temporal argument is found by $\pi-(\frac{\pi}{2}-\vec{\beta}\cdot(-\hat{x}))=\frac{\pi}{2}-\vec{\beta}\cdot\hat{x}$.

The conserved charge \eqref{charge maxwell 4} is rewritten as
\begin{equation}
Q_\laa=-\int_{S^{d-2}}\sqrt{q}\left(\laa\pi^r-\mu\bar{A}_r\right)
\end{equation}
 One must note that $\mu$ transforms like a vector under boosts, for it is the $T$-derivative of a scalar.

\subsection{Finite action and symplectic form}
Here we will show that the symplectic form is finite in dimensions higher than 4. In analogy with mechanical systems, the symplectic 2-form $\Omega$ in field theories is defined from the boundary term of the Lagrangian. For Maxwell theory in Rindler patch, it is
\begin{equation}
\Omega=-\int_I\sqrt{g}\delta\A_{\nu}\delta\F^{\mu\T}+\Omega^{\text{\tiny b'dary}}\,,
\end{equation}
with $\Omega^{\text{\tiny b'dary}}$ being a surface term introduced in \cite{Campiglia:2017mua} for $d=4$ \footnote{It exists also in higher dimensions. We did not need to introduce it for the charges were derived from the action.}. In four dimensions, this is logarithmically divergent, since
\begin{equation}
\Omega=\Omega^{\text{\tiny b'dary}}-\int_I\sqrt{h}d\hat{x}\frac{d\rho}{\rho}\left(\delta A_\rho^{(1)}\delta F^{\rho\T}_{(1)}+
\delta A_\Bi^{(0)}\delta F^{\Bi\T}_{(0)}
\right)+\ord{\rho^{0}}
	\end{equation}
The second term which correponds to magnetic monopoles is  excluded in our boundary condition \eqref{maxwell b.c.}. The first term, however has the form $\int\psi\partial_\T\psi$. If the integration surface is $\Ti=\pi/2$, this term vanishes by antipodal condition \eqref{parity cond1}. This remains true for boosted frames too. Nevertheless, it is not clear if the divergence cancels for arbitrary spacelike surfaces $I$, and we are not aware of any resolution. Similar divergence occurs in computing the on-shell action, where the cancellation around $\Ti=\pi/2$ surface is again ensured by antipodal conditions.

In higher dimensions, $\delta A^{(1)}_{\rho}=0$, and no large $\rho$ divergence appears.

\section{Three dimensions}\label{3d sec}
This section is devoted to three dimensional Maxwell theory. The asymptotic symmetry at null infinity was discussed in \cite{Barnich:2015jua}. The reason for separate consideration of three dimensional case is the simple form of solutions:  dS$_2$ is conformally flat and the solution is a whole set of left- and right-moving scalar modes. For this simplest case, we will translate the boundary conditions to Bondi coordinates $(u,r,\varphi)$.
\subsection{Boundary conditions and solution space}
The boundary conditions \eqref{maxwell b.c.} for $d=3$ become
\footnote{The static Coulomb solution is
$
\F_{tr}=q/r\,.    
$ 
The electric field in hyperbolic coordinates becomes $\F_{\T\rho}=-q\,$.}
\begin{equation}
\F_{a\rho
}\sim \ord{\rho^0}\,,\qquad \F_{bc}\sim \ord{\rho^{0}}\,.
\end{equation}
This boundary condition is realized by following fall-off on the gauge field
\begin{equation}\label{3d falaf}
    \A_{\rho}\sim \ord{\rho^0}\,,\qquad \A_{a}\sim \ord{\rho^0}\,.
\end{equation}
The asymptotic behavior adopted here allows for moving charges in $2+1$ dimensions.
At leading order, $F^{(0)}_{a\rho}=\partial_a A^{(0)}_\rho$. Let us denote $A^{(0)}_\rho$ by $\psi$ for notational harmony with higher dimensions. Its equation of motion is
\begin{equation}\label{3d eq}
    D_aD^a\psi=0\,.
\end{equation}
The differential operator is the Laplacian on $dS_2$, which takes a nicer form in coordinates $x^\pm=\varphi\pm\Ti$. The metric on $dS_2$ is
\begin{equation}
    d\tilde{s}^2=\frac{-d\Ti^2+d\varphi^2}{\sin^2 \Ti}=\frac{\di x^+\di x^-}{\sin^2 \Ti}.
\end{equation}
The field equation\eqref{3d eq} becomes
\begin{equation}
    \partial_+\partial_-\psi=0\,.
\end{equation}
The general solution  with periodic boundary condition $\psi(\Ti,\varphi)=\psi(\Ti,\varphi+2\pi)$ is the following.
\begin{equation}\label{3d Maxwell sol}
    \psi(\Ti,\varphi)=a_0+b_0\Ti+\sum_{n\neq 0}\left(a_n e^{inx^+}+b_ne^{inx^-}\right)
\end{equation}

\subsection{Action principle and charges}
The boundary term with fall-off \eqref{3d falaf} is finite
\begin{eqnarray}
\int_{B}\sqrt{h}\delta A_{a}^{(0)}\partial^a\psi
\end{eqnarray}
Integration by parts and fixing the asymptotic gauge  $D^aA^{(0)}_a=0$ makes the integrand a total divergence. In contrast to higher dimensions, fixing the Lorenz gauge $\nabla_\mu\A^\mu$ is not possible, because it implies either $\psi=0$ or $\A_a\sim\ord{\rho}$. 

The asymptotic gauge fixing leaves residual gauge transformations satisfying $D_aD^a\laa=0$. The conserved charges are obtained by the same method explained before.
\begin{equation}
Q_\laa=\int_{S^1}\sqrt{h}\left(\partial_\T\laa \psi-\laa \partial_\T \psi\right)
\end{equation}

\subsection*{Antipodal condition}
The whole set of solutions \eqref{3d Maxwell sol} are regular at light cone. Nevertheless, we opt to impose conditions \eqref{parity cond1} which include physical solutions.
\begin{equation}
    \psi(\Ti,\varphi)=-\psi(\pi-\Ti,\varphi+\pi)
\end{equation}
The antipodal map $(\Ti,\varphi)\to(\pi-\Ti,\varphi+\pi)$ is equivalent to $x^+\leftrightarrow x^-$. As a result,  \eqref{3d Maxwell sol} is divided into even and odd parts
\begin{subequations}
\begin{align}
\psi(\Ti,\varphi)&=c_0 \Ti+\sum_{n\neq 0} \frac{c_n}{n} e^{in\varphi}\sin n\Ti\,,\qquad c_n=c_{-n}^\ast \qquad \text{odd}\label{odd} \\  
\laa(\Ti,\varphi)&=d_0 +\sum_{n\neq 0} d_n e^{in\varphi}\cos n\Ti\,,\qquad d_n=d_{-n}^\ast    \qquad \text{even}
\end{align}
\end{subequations}
By this boundary condition, the field strength is obtained by taking a derivative of $\psi$. One can explicitly check that for a boosted electric charge, the gauge field lies in \eqref{odd}.

\subsection*{Charge and boundary conditions at null limit}
Define
\begin{equation}
\bar{ \laa}(\varphi)=\sum_n d_n e^{in\varphi}\qquad\qquad \bar{ \psi}(\varphi)=\sum_n c_n e^{in\varphi}
\end{equation}
 Close to the future null infinity  at $\Ti=0$, the fields behave as follows
\begin{subequations}
\begin{align}
\psi&=\bar{ \psi}(\varphi)\Ti+\ord{\Ti^3}\\
\laa&=\bar{ \laa}(\varphi)+\ord{\Ti^2}
\end{align}
\end{subequations}

At null infinity, only the second term of the charge remains non-vanishing, so the charge is
\begin{equation}
Q_\laa=\int_{S^1}\sqrt{q}\bar{\laa}\bar{\psi}
\end{equation}

To make contact with results \cite{Barnich:2015jua} let us rewrite the boundary conditions near null infinity $(\rho\to\infty,T\to 0)$ in Bondi coordinates $(u,r,x^A)$. The coordinates are related by $u=-\rho  \Ti/2$ and $r=\rho / \Ti$. Expanding the asymptotic gauge $D^aA^{(0)}_{a}=0$ we have
\begin{equation}
\sin^2\Ti\left(\partial_\T A_\T^{(0)}-\partial_\varphi A_\varphi^{(0)}\right)=0\,.
\end{equation}
This condition can be solved by introducing a scalar $\alpha(\Ti,\varphi)$\,.
\begin{equation}
A_\T^{(0)}=\partial_\varphi\alpha\qquad\qquad 
A_\varphi^{(0)}=\partial_\T\alpha
\end{equation}
We have to assign a fall-off for $\alpha$ around $\Ti=0$. Analyzing the dipole solutions, the appropriate condition is $\alpha(\Ti,\varphi)=\bar{ \alpha}\Ti+\ord{\Ti^2}$. Now we can find $\A_u,\,\A_r$ and $\A_{\varphi}$ at leading order.
\begin{align}\label{3d null bc}
A_\varphi&=\bar{ \alpha}(\varphi)+\partial_\varphi\bar{ \laa}(\varphi)+\ord{r^{-1}}\\
A_u&=\bar{\psi}(\varphi)+\ord{r^{-1}}\\
A_r&=-\frac{u}{r}\bar{\psi}(\varphi)+\ord{r^{-2}}
\end{align}

These results must not be interpreted as null infinity boundary conditions. To account for electromagnetic radiation, there should exist one arbitrary function both of $u$ and $\varphi$, corresponding to the single helicity state of photon in three dimensions. Nontheless, \eqref{3d null bc} provides a consistent boundary condition at past of future null infinity, where the radiation has not yet started.

\section{Discussion}

In this paper, we considered asymptotic symmetries of Maxwell theory in three and higher dimension at spatial infinity. We tried to bypass standard methods for computing surface charges, by making the action principle well-defined, applying a gauge transformation on it, and interpreting the resulting conserved quantity as the charge. This work excludes magnetic charges to avoid technical difficulties, although they are discussed in various four dimensional treatments.

We showed that regularity of field strength tensor at light cone implies a certain antipodal condition on de Sitter space in four and higher dimensions, which was familiar in dS/CFT context. In addition, the charges depend on the scalar field $\psi$ on de Sitter space in all dimensions. It is interesting if  dS/CFT quantum considerations  applied to $\psi$ have implications on Maxwell theory.

In three dimension, the solution space is more transparent as the asymptotic de Sitter space is conformally flat. The light cone regularity argument does not work in three dimension, although it is satisfied by the solution for moving electric charges. For this simple model, we could solve the gauge condition and translate the boundary conditions into Bondi coordinates which are better suited for null infinity discussions.

As an interesting generalization, note that in three dimensions, non-trivial vorticity for gauge field is possible. Gauge transformations considered here are regular, so preserve vorticity. Addition of singular gauge transformations which lead to vorticity might lead to an unexpected relation with electric charges considered here; as is the case in four dimensions \cite{Hosseinzadeh:2018dkh,Freidel:2018fsk}.

Finally, we compared our treatment with Hamiltonian formulations of the theory.  Symplectic form and on-shell action are finite in $d>4$ and their divergences $d=3,4$ cancel in inertial frames by virtue of parity conditions. Nonetheless, cancellation in arbitrary slice of asymptotic de Sitter space remains elusive.

\acknowledgments
The author is greatly indebted to Shahin Sheikh-Jabbari and Hamid Afshar for illuminating discussions. He also thanks Miguel Campiglia, Vahid Hoseinzadeh, Roberto Oliveri and especially Ali Seraj for helpful comments on the draft.
\bibliographystyle{fullsort.bst}
 
\bibliography{review} 

\end{document}